\documentclass[11pt,twoside]{article}

\usepackage{amsmath}
\usepackage{amssymb}
\usepackage{cite}
\usepackage{ccaption}
\usepackage{hyperref}
\usepackage[pdftex]{graphicx}
\captiondelim{.}

\begin{document}
\newcommand{\pst}{\hspace*{1.5em}}

\newcommand{\rigmark}{\em Journal of Russian Laser Research}
\newcommand{\lemark}{\em Volume 30, Number 5, 2009}

\newcommand{\be}{\begin{equation}}
\newcommand{\ee}{\end{equation}}
\newcommand{\bm}{\boldmath}
\newcommand{\ds}{\displaystyle}
\newcommand{\bea}{\begin{eqnarray}}
\newcommand{\eea}{\end{eqnarray}}
\newcommand{\ba}{\begin{array}}
\newcommand{\ea}{\end{array}}
\newcommand{\arcsinh}{\mathop{\rm arcsinh}\nolimits}
\newcommand{\arctanh}{\mathop{\rm arctanh}\nolimits}
\newcommand{\bc}{\begin{center}}
\newcommand{\ec}{\end{center}}

\thispagestyle{plain}

\label{sh}


\begin{center} {\Large \bf
\begin{tabular}{c}
Contextuality and probability
\\[-1mm]
representation of quantum states
\end{tabular}
 } \end{center}

\bigskip

\bigskip

\begin{center} {\bf
A.A. Strakhov$^1$ and V.I. Man'ko$^2$
}\end{center}

\medskip

\begin{center}
{\it
$^1$Moscow Institute of Physics and Technolodgy\\
Institutskii per. 9, Dolgoprudnyi, Moscow Region 141700, Russia

\smallskip

$^2$P.N. Lebedev Physical Institute, Russian Academy of Sciences\\
Leninskii Prospect 53, Moscow 119991, Russia
}
\smallskip

\end{center}

\begin{abstract}\noindent
The contextuality and noncontextuality notions are considered in framework of probability representation of quantum states. Example of qutrit states and violation of the noncontextuality inequalities are presented by using the spin tomogram and tomographic symbols of the observables.
\end{abstract}

\medskip


\section{Introduction}
\pst
The quantum correlations of composite systems can be explicitly demonstrated if the systems are in entangled states. For example, for the qubits the violation of Bell inequalities~\cite{1} provides the essential difference of the correlations in separable and nonseparable states. For systems which are not composed ones the example of quantum correlations is given by violation of other kinds of inequalities~\cite{2,3,4,5,6,7,8}.

These inequalities correspond to relation of marginal probability distributions of subsets of random variables with joint probability distributions creating the marginals. It is known that there exist the distributions of a few random variables which can not be obtained as marginals expressed in terms of the joint probability distributions of many random variables. Such relations of the distributions are usually called "contextuality"~\cite{2,9}. If the distributions of a few random variables can be obtained as marginals of the joint probability distribution one says that we have noncontextuality situation. Examples of contextuality, corresponding to random variables are known for the case of quantum states and related to the state properties of the set of quantum observables~\cite{3,4}.

Recently~\cite{10,11} new formulation of quantum mechanics was suggested. In this formulation quantum states are identified with fair probability distributions (called usually quantum tomograms). The quantum observables are identified with the tomographic symbols of the corresponding hermitian operators. In the tomographic probability representation the formalism of considering the statistics of classical and quantum observables has the common basis which is related to standard notions and tools of the probability theory.

The aim of our work is to consider the contextuality property of quantum measurements in the probability representation of quantum states. We reformulate some of the results~\cite{4,7} identifying both the states and normalized projectors with tomographic probability distributions.

The paper is organized as follows. In section 2 the entropic noncontextual inequalities are discussed. In section 3 the spin-tomogram is reviewed. In section 4 some of the noncontextual inequalities are discussed in framework of spin-tomographic probability representation of quantum mechanics. In section 5 another view on joint probabilities is presented and in section 6 unitary tomograms are discussed. Section 7 contain summary of the work. In Appendix some details of used formulas are given.

\section{Noncotextuality inequalities in quantum systems}
\pst
First of all, let us review some facts from contextuality approach. Noncontextuality definition reads: the system with $N$ random variables is noncontextual, if there exists a joint probability distribution $P(A_1,A_2,\ldots ,A_N)$. Otherwise, it is contextual. From general considerations and existence of this probability distribution, there can be derived several inequalities, which violation shows contextual character of systems. These inequalities can be applied to quantum systems.

We assume that for random variables considered below there exists joint probability distribution and the means, variances and covariances are calculated in view of standard formalism of probability theory.

Klyachko-Can-Binicio$\breve{\text{g}}$lu-Shumovsky~\cite{3} inequality works for 5 random variables with outcomes 1 and -1 and it reads:
\be
\langle A_1A_2\rangle+\langle A_2A_3\rangle+\langle A_3A_4\rangle+\langle A_4A_5\rangle+\langle A_5A_1\rangle\geq-3.
\label{eq}
\ee
For quantum systems in next example means, variances and covariances are calculated by using formalism of density operator. Thus, in three-dimensional Hilbert space and specially taken 5 real vectors and a real state, this inequality is sometimes called pentagram inequality and looks like~\cite{3}:
\be
\langle S_{\vec{l}_1}^2\rangle_{\psi}+\langle S_{\vec{l}_2}^2\rangle_{\psi}+\langle S_{\vec{l}_3}^2\rangle_{\psi}+\langle S_{\vec{l}_4}^2\rangle_{\psi}+\langle S_{\vec{l}_5}^2\rangle_{\psi}\geq3,
\ee
where the vectors $\vec{l}_i\perp\vec{l}_{i+1}$ for $i=1,\ldots,4$ and $\vec{l}_5\perp\vec{l}_{1}$. The operator $S^2_{\vec{l}_k}=(\hat{\vec{J}},\vec{l}_k)^2,$ where the components of vector $\hat{\vec{J}}$ are generators of rotation group.

The inequality (\ref{eq}) was generalized~\cite{5} for $N$ measurements:
\be
\langle \chi\rangle=\sum\limits_{i=1}^{N-1}\langle A_i A_{i+1}\rangle+(-1)^{N-1}\langle A_N A_{1}\rangle\geq-(N-2).
\ee
In case of odd $N\geq 4$ the minimal possible quantum mean value $\langle \chi\rangle$ for three-dimensional system is $\Omega_N=-\frac {3N\cos{(\pi/N)}-N}{1+\cos{(\pi/N)}}.\;$In case of even $N\geq 4$ minimal $\langle \chi\rangle$ is $\Xi_N=-N\cos{(\pi/N)}$ for four-dimensional system and $(-1+\Xi_{N-1})$ for three-dimensional system with observables, that can be proportional to identity or $A_i=\pm A_{i+1}$. These quantum mean values violate the above inequality.

There exists another inequality called Peres-Mermin inequality and it is based on relations between 9 random variables $A,B,C,a,b,c,\alpha,\beta$ and $\gamma$ with outcomes 1 and -1~\cite{12,13,14}:
\be
\langle \chi\rangle=\langle ABC\rangle+\langle bca\rangle+\langle \gamma\alpha\beta\rangle+\langle A\alpha a\rangle+\langle bB\beta\rangle-\langle \gamma cC\rangle\leq 4.
\ee
However, this inequality violates for a quantum system known as Peres-Mermin square:
$$A=\sigma_{x}^1 \otimes I^{2},\quad B=I^{1} \otimes \sigma_{x}^2,\quad C=\sigma_{x}^1 \otimes \sigma_{x}^2,$$
$$a=I^{1} \otimes \sigma_{y}^2,\quad b=\sigma_{y}^1 \otimes I^2,\quad c=\sigma_{y}^1 \otimes \sigma_{y}^2,$$
$$\alpha=\sigma_{x}^1 \otimes \sigma_{y}^2,\quad \beta=\sigma_{y}^1 \otimes \sigma_{x}^2,\quad \gamma=\sigma_{z}^1 \otimes \sigma_{z}^2,$$
where upper index represents two-dimensional subsystem of a four-dimensional system and I - identity operator, $\sigma_{i}$ - Pauli matrices. For an arbitrary state mean value $\langle \chi\rangle$ is equal to 6.

Now let us consider entropic inequality. It is based on Shannon entropy. Here we will give definitions and important properties of this entropy. Shannon entropy for a discrete random variable defines as follows:
$$H(A)=-\sum_{a}P(A=a)\log_2{P(A=a)},$$ where $P(A=a)$ is a probability of an elementary outcome. Then we can define entropy for a joint probability distribution and conditional entropy:
$$H(AB)=-\sum_{a,b}P(A=a,B=b)\log_2{P(A=a,B=b)};\; H(A|B)=\sum_{b}P(B=b)H(A|B=b),$$
where $H(A|B=b)=-\sum_{a}P(A=a|B=b)\log_2{P(A=a|B=b)}$. According to these definitions, there can be simply derived these properties:
$$H(AB)=H(A|B)+H(B);\quad H(A|B)\leq H(A)\leq H(AB).$$ Based on this two properties, the following statement can be proved (for example, using mathematical induction) - the system of $n$ random variables is noncontextual implies that the following inequality takes place:
\be
H(A_1|A_n)\leq H(A_1|A_2)+H(A_2|A_3)+\ldots+H(A_{n-1}|A_n).
\ee
Now we can apply this inequality for a special quantum system - 5 projectors $\hat{R}_i=|A_i\rangle\langle A_i|$ in three-dimensional Hilbert space, where $|A_i\rangle\perp|A_{i+1}\rangle,\;|A_{6}\rangle=|A_{1}\rangle$ for $i=1,2,\ldots,5.$ Every projector has two eigenvalues: 0 and 1. Thus, we can connect with each projector a probability distribution with two elementary outcomes: 0 and 1. If we take an arbitrary state $|\Psi\rangle$, this probability distribution for the projector $\hat{R}_i$ will be:
$$P_i=
\begin{cases}
|\langle A_i|\Psi\rangle|^2,&\text{if the outcome is 1}\\
1-|\langle A_i|\Psi\rangle|^2,&\text{if the outcome is 0}
\end{cases}$$ We can also construct joint probability distributions $P_{i,i+1}$ for neighboring projectors due to $|A_i\rangle\perp|A_{i+1}\rangle,\;|A_{6}\rangle=|A_{1}\rangle$ for $i=1,2,\ldots,5:$
\be
\label{eq6}
P_{i,i+1}=
\begin{cases}
1-|\langle A_i|\Psi\rangle|^2-|\langle A_{i+1}|\Psi\rangle|^2,&\text{if the outcome is 00}\\
|\langle A_i|\Psi\rangle|^2,&\text{if the outcome is 10}\\
|\langle A_{i+1}|\Psi\rangle|^2,&\text{if the outcome is 01}\\
0,&\text{if the outcome is 11}
\end{cases}
\ee
After definition of these probability distributions we are able to calculate conditional entropies $H(A_i|A_{i+1})=H(A_i,A_{i+1})-H(A_{i+1})$ and to check entropy inequality. The violation of this inequality was discovered on the following vectors~\cite{4}:
$$|\Psi\rangle=
\begin{pmatrix}
\sin{\theta} \\
\cos{\theta} \\
0
\end{pmatrix};\;
|A_1\rangle=
\begin{pmatrix}
\frac{\sqrt{\cos{2\phi}}}{\sqrt{2}\cos{\phi}} \\
\frac{\tan{\phi}}{\sqrt{2}} \\
\frac{1}{\sqrt{2}}
\end{pmatrix};\;
|A_2\rangle=
\begin{pmatrix}
0 \\
\cos{\phi} \\
-\sin{\phi}
\end{pmatrix};\;$$
$$
|A_3\rangle=
\begin{pmatrix}
1 \\
0 \\
0
\end{pmatrix};\;
|A_4\rangle=
\begin{pmatrix}
0 \\
\cos{\phi} \\
\sin{\phi}
\end{pmatrix};\;
|A_5\rangle=(|A_1\rangle\times |A_4\rangle)/\||A_1\rangle\times |A_4\rangle\|,$$
where the angles were: $\theta=0.2366,\; \phi=0.1698.$

This inequality will be studied further. In section 4 we will show, how to calculate probabilities, introduced above, using the tomographic picture of quantum mechanics.

\section{Spin tomograms}
\pst
In this section we will review construction of spin tomograms according to the general scheme of tomographic symbols~\cite{15}. In general case, we consider a Hilbert space $H$ with an operator $\hat{A}$ acting on it. Let us suppose that we have operator function $\hat{U}(x)$ depending on $x=(x_1,x_2,...,x_n)$-set of n parameters. Then, we can construct a $c$-number function $$f_{\hat{A}}(x)=Tr[\hat{A}\hat{U}(x)].$$ Let us suppose the relation has an inverse, i.e., there exists a set of operators $\hat{D}(x)$ such that $$\hat{A}=\int f_{\hat{A}}(x)\hat{D}(x)dx.$$ Function $f_{\hat{A}}(x)$ is called tomographic symbol of operator $\hat{A}$ and operator $\hat{U}(x)$ called dequantizer, $\hat{D}(x)$ - quantizer.

Let us represent an arbitrary operator $\hat{A}^{(j)}$ by the matrix $$\hat{A}^{(j)}_{mm'}=\langle jm|\hat{A}^{(j)}|jm'\rangle,\quad m=-j,-j+1,...,j,$$ where $$\hat{j}_3|jm\rangle=m|jm\rangle;\quad \hat{j}^2|jm\rangle=j(j+1)|jm\rangle,$$ and $$\hat{A}^{(j)}=\sum^{j}_{m=-j}\sum^{j}_{m'=-j}\hat{A}^{(j)}_{mm'}|jm\rangle\langle jm'|.$$ Now let us introduce a finite rotation operator $U$ of $SU(2)$ group. Then we can determine dequantizer as follows
$$
\hat{\tilde{U}}(x)=U|jm_1\rangle\langle jm_1|U^{\dag}.
$$
Rotation transform $U$ depends on Euler angles $\alpha$, $\beta$,$\gamma$, and matrix elements of this transform are
\be
D^{(j)}_{m'm}(\alpha,\beta,\gamma)=e^{\imath m'\gamma}d^{(j)}_{m'm}(\beta)e^{\imath m\alpha},
\label{eq7}
\ee
where $$d^{(j)}_{m'm}(\beta)=\left [\frac{(j+m')!(j-m')!}{(j+m)!(j-m)!}\ \right ]^{1/2}\left ( \cos{\frac{\beta}{2}}\right )^{m'+m}\left ( \sin{\frac{\beta}{2}}\right )^{m'-m}P^{(m'-m,m'+m)}_{j-m'}(\cos\beta),$$ where $P^{(a,b)}_{n}(x)$ is Jacobi polynomial. Then tomographic symbol $\omega(x)=Tr[\hat{A}\hat{\tilde{U}}(x)]$ will be:
\be
\omega(m_1,\alpha,\beta)=\sum^{j}_{m'_1=-j}\sum^{j}_{m'_2=-j}D^{(j)}_{m_1m'_1}(\alpha,\beta,\gamma)A^{(j)}_{m'_1m'_2}D^{(j)*}_{m_1m'_2}(\alpha,\beta,\gamma).
\ee

The inverse map between $\hat{A}^{(j)}$ and $\omega(x)$ can be achieved with following quantizer $\hat{D}(x)$:$$\hat{D}(x)=\frac {1}{8\pi^2}\hat{B}^{(j)}_{m_1}(\alpha,\beta),$$ $$\hat{B}^{(j)}_{m_1}(\alpha,\beta)=(-1)^{m_1}\sum^{2j}_{j_3=0}\begin{pmatrix} j & j & j_3 \\ m_1 & -m_1 & 0 \end{pmatrix}\hat{A}^{(j_3)}_{j}(\alpha,\beta),$$ $$\hat{A}^{(j_3)}_{j}(\alpha,\beta)=(2j_3+1)^2\sum^{j}_{m'_1=-j}\sum^{j}_{m'_2=-j}| jm'_1\rangle \Phi^{(j_3)}_{jm'_1m'_2}(\alpha,\beta)\langle jm'_2|,$$ $$\Phi^{(j_3)}_{jm'_1m'_2}(\alpha,\beta)=(-1)^{m'_2}\sum^{j_3}_{m_3=-j_3}D^{(j_3)}_{0m_3}(\alpha,\beta,\gamma)\begin{pmatrix} j & j & j_3 \\ m'_1 & -m'_2 & m_3 \end{pmatrix}.$$ Then, for $\hat{A}^{(j)}$ we will have:
\be
\hat{A}^{(j)}=\sum^{j}_{m_1=-j}\int \frac{d\Omega}{8{\pi}^2}\omega(m_1,\alpha,\beta)\hat{B}^{(j)}_{m_1}(\alpha,\beta);\quad\quad \int d\Omega=\int \limits_{0}^{2\pi}d\alpha\int \limits_{0}^{2\pi}d\gamma\int \limits_{0}^{\pi}\sin{\beta}d\beta.
\ee
We can also construct a dual tomographic symbol $\omega^{d}(x)$: $\omega^{d}(x)=Tr[\hat{A}\hat{D}(x)].$ It will allow us to calculate average value of an arbitrary observable $\hat{A}$:
\be
\label{eq10}
\langle A\rangle=Tr[\hat{\rho}\hat{A}]=\int \omega_{\hat{\rho}}(x)\omega^{d}_{\hat{A}}(x)dx.
\ee

\section{Entropic inequality in framework of spin tomography}
\pst
In this section we will apply formalism of spin tomograms to calculate probabilities and, hence, Shannon entropies. As it was showed (\ref{eq6}), we need to calculate probabilities $|\langle A_i|\Psi\rangle|^2$. Using equation (\ref{eq10}) for mean value of an observable, we can achieve: $$|\langle A_i|\Psi\rangle|^2=\langle \Psi|\hat{R}_i|\Psi\rangle=Tr[\hat{\rho}\hat{R}_i]=\int \omega_{\hat{\rho}}(x)\omega^{d}_{\hat{R}_i}(x)dx,$$ where $\hat{\rho}=|\Psi\rangle\langle \Psi|$ and $\hat{R}_i=|A_i\rangle \langle A_i|$. Let us show, that for three-dimensional Hilbert space and its real vectors $|\Psi\rangle$ and $|A_i\rangle$ we will get the same result for calculating $|\langle A_i|\Psi\rangle|^2$, as in traditional quantum mechanics: $|\langle A_i|\Psi\rangle|^2=(A_{i1}\Psi_1+A_{i2}\Psi_2+A_{i3}\Psi_3)^2$. We will show this for state $|\Psi\rangle=\begin{pmatrix} \sin{\theta} \\ \cos{\theta} \\ 0\end{pmatrix}$ and an arbitrary real vector $|A\rangle=\begin{pmatrix} A_1 \\ A_0 \\ A_{-1}\end{pmatrix}$. First of all, we will derive $\omega_{\hat{\rho}}(x)$:
$$\omega_{\hat{\rho}}(m,\alpha,\beta)=
  \sum^{1}_{m'_1=-1}\sum^{1}_{m'_2=-1}D^{(1)}_{mm'_1}(\alpha,\beta,\gamma)\rho_{m'_1m'_2}D^{(1)*}_{mm'_2}(\alpha,\beta,\gamma)=$$
$$=\sum^{1}_{m'_1=-1}\sum^{1}_{m'_2=-1}d^{(1)}_{mm'_1}(\beta)
  \rho_{m'_1m'_2}d^{(1)*}_{mm'_2}(\beta)e^{\imath (m'_1-m'_2)\alpha};$$
$$\rho=\begin{pmatrix}
{\sin}^2\theta & \frac{\sin{2\theta}}{2} & 0 \\
\frac{\sin{2\theta}}{2} & {\cos}^2\theta & 0 \\
0 & 0 & 0
\end{pmatrix}; \quad\quad
d^{(1)}(\beta)=\begin{pmatrix}
\frac{1}{2}(1+\cos{\beta}) & \frac{1}{\sqrt{2}}\sin{\beta} & \frac{1}{2}(1-\cos{\beta}) \\
-\frac{1}{\sqrt{2}}\sin{\beta} & \cos{\beta} & \frac{1}{\sqrt{2}}\sin{\beta} \\
\frac{1}{2}(1-\cos{\beta}) & -\frac{1}{\sqrt{2}}\sin{\beta} & \frac{1}{2}(1+\cos{\beta})
\end{pmatrix}.$$
Eventually, spin tomogram will be:
\bea
\omega_{\hat{\rho}}(1,\alpha,\beta)
  =\frac{1}{4}(1+\cos{\beta})^2{\sin}^2{\theta}+\frac{1}{2}{\sin}^2{\beta}{\cos}^2{\theta}+
  \frac{1}{2\sqrt{2}}\sin{\beta}(1+\cos{\beta})\sin{2\theta}\cos{\alpha}; \\
\omega_{\hat{\rho}}(0,\alpha,\beta)=\frac{1}{2}{\sin}^2{\beta}{\sin}^2{\theta}+
  {\cos}^2{\beta}{\cos}^2{\theta}-\frac{1}{2\sqrt{2}}\sin{2\beta}\sin{2\theta}\cos{\alpha}; \\
\omega_{\hat{\rho}}(-1,\alpha,\beta)=\frac{1}{4}(1-\cos{\beta})^2{\sin}^2{\theta}+
  \frac{1}{2}{\sin}^2{\beta}{\cos}^2{\theta}-\frac{1}{2\sqrt{2}}\sin{\beta}(1-\cos{\beta})\sin{2\theta}\cos{\alpha}.
\eea
Now let us derive $\omega^{d}_{\hat{R}}(x)$:
\begin{multline*}
\omega^{d}_{\hat{R}}(m,\alpha,\beta)=\frac{(-1)^m}{8{\pi}^2}\sum^{2}_{j_3=0}
\begin{pmatrix}
1 & 1 & j_3 \\
m & -m & 0 \\
\end{pmatrix}
  (2j_3+1)^2\sum^{1}_{m'_1=-1}\sum^{1}_{m'_2=-1}A_{m'_1}A_{m'_2}(-1)^{m'_2}\times\\
   \times\sum^{j_3}_{m_3=-j_3}D^{(j_3)}_{0m_3}(\alpha,\beta,\gamma)
\begin{pmatrix}
1 & 1 & j_3 \\
m'_1 & -m'_2 & m_3 \\
\end{pmatrix}
\end{multline*}

After numerous calculations we can achieve:
\begin{eqnarray}
\omega^{d}_{\hat{R}}(m,\alpha,\beta)=\frac{1}{24{\pi}^2}(A_1^2+A_{-1}^2+A_0^2)+
\frac{3m}{16{\pi}^2}\bigl[\sqrt{2}(A_1A_0+A_{-1}A_0)\sin{\beta}\cos{\alpha}+\cos{\beta}(A_1^2-A_{-1}^2)\bigr]+
    \nonumber \\
+\frac{5(3m^2-2)}{48{\pi}^2}\bigl[3A_1A_{-1}\cos{2\alpha}{\sin}^2{\beta}+
   \frac{3}{\sqrt{2}}(A_1A_0-A_{-1}A_0)\sin{2\beta}\cos{\alpha}+ \nonumber\\
\frac{1}{2}(A_1^2+A_{-1}^2-2A_0^2)(3{\cos}^2{\beta}-1)\bigr].
\end{eqnarray}
So, finally, we need to calculate following integral:
$$|\langle A|\Psi\rangle|^2=\langle \hat{R}\rangle=\sum^{1}_{m=-1}\int \limits_{0}^{2\pi}d\alpha\int \limits_{0}^{2\pi}d\gamma\int \limits_{0}^{\pi}\omega_{\hat{\rho}}\omega^{d}_{\hat{R}}\sin{\beta}d\beta=A_1^2{\sin}^2{\theta}+2A_1A_0\sin{\theta}\cos{\theta}+A_0^2{\cos}^2{\theta}=$$
\be
\label{eq15}
=(A_1\sin{\theta}+A_0\cos{\theta})^2
\ee
As we can see, it is equal to what we calculate using standard scalar product.
\section{Spin tomograms and joint probability}
\pst
One can see that the vectors $|A_k\rangle$ and $|\psi\rangle$(which were introduced in sec. 2) can be written in the form $$|A_k\rangle=U_k|A_3\rangle,\;k=1,2,4,5;\quad |\psi\rangle=U_{\psi}|A_3\rangle.$$
The matrices $U_k$ and $U_\psi$ are the unitary matrices:
$$
U_1=
\begin{pmatrix}
\frac {\sqrt{\cos2\phi}}{\sqrt{2}\cos\phi} & -\frac {\cos\phi}{\sqrt{3\cos^2\phi-1}} & \tan\phi\sqrt{\frac {\cos2\phi}{3\cos2\phi+1}} \\
\frac {\tan\phi}{\sqrt{2}} & 0 & -\frac {2\cos\phi-\tan\phi\sin\phi}{\sqrt{3\cos2\phi+1}} \\
\frac {1}{\sqrt 2} & \frac {\sqrt{\cos 2\phi}}{\sqrt{3\cos^2\phi-1}} & \frac {\sin\phi}{\sqrt{3\cos 2\phi+1}}
\end{pmatrix};\quad
U_2=
\begin{pmatrix}
0 & 0 & 1 \\
\cos\phi & \sin\phi & 0 \\
-\sin\phi & \cos\phi & 0
\end{pmatrix};$$
$$U_4=
\begin{pmatrix}
0 & 0 & 1 \\
\cos\phi & -\sin\phi & 0 \\
\sin\phi & \cos\phi & 0
\end{pmatrix};\quad
U_5=
\begin{pmatrix}
-\frac {\sqrt{\cos2\phi}}{\sqrt{2}\cos\phi} & \frac {\cos\phi}{\sqrt{3\cos^2\phi-1}} & -\tan\phi\sqrt{\frac {\cos2\phi}{3\cos2\phi+1}} \\
-\frac {\tan\phi}{\sqrt{2}} & 0 & \frac {2\cos\phi-\tan\phi\sin\phi}{\sqrt{3\cos2\phi+1}} \\
\frac {1}{\sqrt 2} & \frac {\sqrt{\cos 2\phi}}{\sqrt{3\cos^2\phi-1}} & \frac {\sin\phi}{\sqrt{3\cos 2\phi+1}}
\end{pmatrix};$$
$$U_{\psi}=
\begin{pmatrix}
\sin\theta & -\cos\theta & 0 \\
\cos\theta & \sin\theta & 0 \\
0 & 0 & 1
\end{pmatrix};
$$
These matrices contain as their columns the normalized eigenvectors of five projectors $\hat{P_k}=|A_k\rangle \langle A_k|, \; k=1,2,4,5,\psi.$ These projectors can be interpreted as density operators corresponding to the "pure states" $|A_k\rangle$ and $|\psi\rangle.$ According to the general construction, these density operators provide the state tomograms which are analogs of qutrit tomogram, given in the form:
\be
\omega_k(m,\overrightarrow{n})=\langle m|U\hat{P_k} U^\dag|m\rangle=\langle m|UU_k\hat{P_3} U_k^\dag U^\dag|m\rangle,\; k=1,2,4,5,\psi;
\ee
$$\hat{P_3}=
\begin{pmatrix}
1 & 0 & 0 \\
0 & 0 & 0 \\
0 & 0 & 0
\end{pmatrix}.$$
There matrix $U$ is the matrix of irreducable representation of rotation group $SU(2)$ and it is expressed in terms of Euler angles by equation (\ref{eq7}). The following tomograms are presented in Appendix. Figure \ref{plot1} shows graphs of tomograms for $P_{i},$ where $i=1,2,4,5,\psi$ and $\phi=0.7,\; \theta=0.5$(second and fourth are green and red and occupy the same area). The tomographic symbol for $i=3$ is represented by a curve.

\begin{figure}[ht]
\bc
\includegraphics[width=10cm]{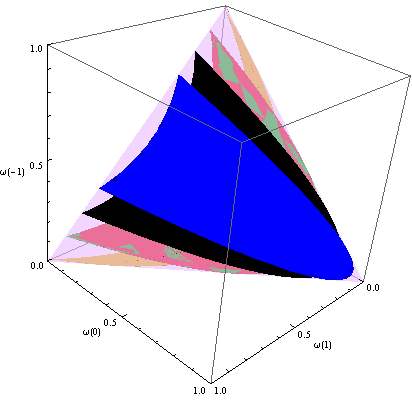}
\ec
\vspace{-4mm}
\caption{ A set of spin tomograms for the state $|\psi\rangle$ and all projectors accept third}
\label{plot1}
\end{figure}

The fidelities providing the probability distributions are given in terms of tomograms in the following form~\cite{16}:
\be
Tr[\hat{P_k}\hat{P_{\psi}}]=(2j+1)\sum_{m=-j}^{j}\int\limits_{S^2}
  \frac{d\overrightarrow{n}}{4\pi}[\omega_k(m,\overrightarrow{n})-
  \frac{1}{2}\omega_k(m+1,\overrightarrow{n})-
  \frac{1}{2}\omega_k(m-1,\overrightarrow{n})]\omega_{\psi}(m,\overrightarrow{n}),
\ee
where $j=1;\;m=-1,0,1$ and integration is produced over the sphere: $\int\limits_{S^2}\frac{d\overrightarrow{n}}{4\pi}=\int\limits_{0}^{2\pi}\int\limits_{0}^{\pi}d\varphi \sin\theta d\theta; \quad \overrightarrow{n}=\begin{pmatrix}
\cos\varphi\sin\theta \\
\sin\varphi\sin\theta \\
\cos\theta
\end{pmatrix}.$
Members $\omega_k(-j-1,\overrightarrow{n})$ and $\omega_k(j+1,\overrightarrow{n})$ of the sum above are zeros.

Calculations give the same result as in equation (\ref{eq15}).
\section{Unitary tomogram}
\begin{figure}[ht]
\bc \includegraphics[width=10cm]{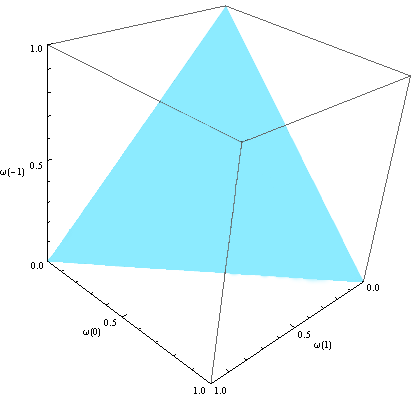}
\ec
\vspace{-4mm}
\caption{ A surface, corresponding to the unitary tomogram}
\label{plot2}
\end{figure}

\pst
Now we will introduce more common, unitary tomogram, based on $U(n)$ group~\cite{17}. We will apply it to qutrit system. The parametrization $U(n)$ is taken from~\cite{18}:
$$
A_n\in U(n);\quad O_n\in O(n);\quad A_3=d_3 O_3 d_2^1 O_2^1 d_1^2;\quad J_{i,i+1}=\begin{pmatrix} I_{i-1} & 0 & 0 & 0 \\ 0 & \cos{\theta_i} & -\sin{\theta_i} & 0 \\ 0 & \sin{\theta_i} & \cos{\theta_i} & 0 \\ 0 & 0 & 0 & I_{n-i-1}\end{pmatrix};
$$
$$
d_{n-k}^k=\begin{pmatrix} I_k & 0 \\ 0 & \operatorname{diag}(e^{\imath\varphi_1},...,e^{\imath\varphi_{n-k}})
\end{pmatrix};\quad O^k_{n-k}=\begin{pmatrix} I_k & 0 \\ 0 & O_{n-k}
\end{pmatrix};\quad O_n=J_{n-1,n}J_{n-2,n-1}...J_{1,2};
$$
$$d_3=\operatorname{diag}(e^{\imath\varphi_1},e^{\imath\varphi_2},e^{\imath\varphi_3});\quad d_2^1=\operatorname{diag}(1,e^{\imath\varphi_4},e^{\imath\varphi_5});\quad d_1^2=\operatorname{diag}(1,1,e^{\imath\varphi_6});\quad \varphi_i\in [0;2\pi)
$$
$$O_3=J_{2,3}J_{1,2}=
\begin{pmatrix}
1 & 0 & 0 \\
0 & \cos{\theta_2} & -\sin{\theta_2} \\
0 & \sin{\theta_2} & \cos{\theta_2}
\end{pmatrix}
\begin{pmatrix}
\cos{\theta_1} & -\sin{\theta_1} & 0 \\
\sin{\theta_1} & \cos{\theta_1} & 0 \\
0 & 0 & 1
\end{pmatrix}=$$
$$=\begin{pmatrix}
\cos{\theta_1} & -\sin{\theta_1} & 0 \\
\cos{\theta_2}\sin{\theta_1} & \cos{\theta_2}\cos{\theta_1} & -\sin{\theta_2} \\
\sin{\theta_2}\sin{\theta_1} & \sin{\theta_2}\cos{\theta_1} & \cos{\theta_2}
\end{pmatrix};$$
$$O_2^1=\begin{pmatrix} I_1 & 0 \\ 0 & O_2 \end{pmatrix}=
\begin{pmatrix}
1 & 0 & 0 \\
0 & \cos{\theta_3} & -\sin{\theta_3} \\
0 & \sin{\theta_3} & \cos{\theta_3}
\end{pmatrix};\quad \theta_j\in [0;\frac {\pi}{2}]$$
If we now calculate the tomogram $\omega(m,A_3)=\langle m|A_3
\begin{pmatrix}
1 & 0 & 0 \\
0 & 0 & 0 \\
0 & 0 & 0
\end{pmatrix}A_3^\dag|m\rangle$, we will achieve:
$\begin{matrix}\omega(1)=\cos^2\theta_1 \\
\omega(0)=\cos^2\theta_2\sin^2\theta_1 \\
\omega(-1)=\sin^2\theta_1\sin^2\theta_2\end{matrix};\\
\theta_1,\theta_2\in [0;\frac {\pi}{2}].$
If one builds a graph of this 2-dimensional surface, it will occupy the whole simplex(see Figure \ref{plot2}).

\section{Summary}
\pst
To resume we point out the main results of our work. The problem of noncontextuality for measuring observables in three-dimensional Hilbert space discussed in~\cite{4,7} is studied in spin-tomographic probability representation of quantum mechanics. The example of five observables, which in the case of the projectors are standard probability distribution, are mapped onto the corresponding regions on the two-dimensional simplex. The probability distributions obtained in ~\cite{4,7} by using standard Born rule are expressed in terms of the tomographic probability distributions related to the projectors(observables) and the tomogram of the state when the observables are measured. The mutual relation between the different probability distribution involved in the calculating violation of inequalities which characterize existence of a joint probability distribution for all measured variables is discussed.

\section{Appendix}
\pst
The result of calculating tomograms for projectors $\hat{R_i}=|A_i\rangle\langle A_i|,\; i=1,...,5$ in section 5:
$$
\omega_1(1,\overrightarrow{n})=
\frac{\frac{\cos2\phi}{\cos^2\phi}(1+\cos\beta)^2+(1-\cos\beta)^2+2\sin^2\beta\tan^2\phi+2\frac{\sqrt{\cos2\phi}}{\cos\phi}\sin^2\beta\cos2\alpha}{8}+$$
$$+\frac{\frac{\sqrt{\cos2\phi}\sin\phi}{\cos^2\phi}\sin\beta(1+\cos\beta)\cos\alpha+\tan\phi\sin\beta(1-\cos\beta)\cos\alpha}{2\sqrt{2}};$$ $$\omega_1(0,\overrightarrow{n})=
\frac {2\cos^2\beta\tan^2\phi+\sin^2\beta+\frac{\cos2\phi}{\cos^2\phi}\sin^2\beta-2\frac {\sqrt{\cos2\phi}}{\cos\phi}\sin^2\beta\cos2\alpha}{4}+\frac {\tan\phi\sin2\beta\cos\alpha}{2\sqrt{2}}-$$
$$-\frac{\frac{\sqrt{\cos2\phi}}{\cos^2\phi}\sin\phi\sin2\beta\cos\alpha}{2\sqrt{2}};$$
$$\omega_1(-1,\overrightarrow{n})=
\frac{\frac{\cos2\phi}{\cos^2\phi}(1-\cos\beta)^2+(1+\cos\beta)^2+2\sin^2\beta\tan^2\phi+2\frac{\sqrt{\cos2\phi}}{\cos\phi}\sin^2\beta\cos2\alpha}{8}-$$
$$-\frac{\frac{\sqrt{\cos2\phi}\sin\phi}{\cos^2\phi}\sin\beta(1-\cos\beta)\cos\alpha+\tan\phi\sin\beta(1+\cos\beta)\cos\alpha}{2\sqrt{2}};$$ $$\omega_2(1,\overrightarrow{n})=
\frac{2\cos^2\phi\sin^2\beta+\sin^2\phi(1-\cos\beta)^2}{4}-\frac{\sin2\phi\sin\beta(1-\cos\beta)\cos\alpha}{2\sqrt{2}};$$
$$\omega_2(0,\overrightarrow{n})=
\frac{2\cos^2\phi\cos^2\beta+\sin^2\phi\sin\beta^2}{2}-\frac{\sin2\phi\sin2\beta\cos\alpha}{2\sqrt{2}};$$
$$\omega_2(-1,\overrightarrow{n})=
\frac{2\cos^2\phi\sin^2\beta+\sin^2\phi(1+\cos\beta)^2}{4}+\frac{\sin2\phi\sin\beta(1+\cos\beta)\cos\alpha}{2\sqrt{2}};$$
$$\omega_3(1,\overrightarrow{n})=\cos^4{\frac{\beta}{2}};$$
$$\omega_3(0,\overrightarrow{n})=\frac{\sin^2{\beta}}{2};$$
$$\omega_3(-1,\overrightarrow{n})=\sin^4{\frac{\beta}{2}};$$
$$\omega_4(1,\overrightarrow{n})=
\frac{2\cos^2\phi\sin^2\beta+\sin^2\phi(1-\cos\beta)^2}{4}+\frac{\sin2\phi\sin\beta(1-\cos\beta)\cos\alpha}{2\sqrt{2}};$$
$$\omega_4(0,\overrightarrow{n})=
\frac{2\cos^2\phi\cos^2\beta+\sin^2\phi\sin\beta^2}{2}+\frac{\sin2\phi\sin2\beta\cos\alpha}{2\sqrt{2}};$$
$$\omega_4(-1,\overrightarrow{n})=
\frac{2\cos^2\phi\sin^2\beta+\sin^2\phi(1+\cos\beta)^2}{4}-\frac{\sin2\phi\sin\beta(1+\cos\beta)\cos\alpha}{2\sqrt{2}};$$
$$
\omega_5(1,\overrightarrow{n})=
\frac{\frac{\cos2\phi}{\cos^2\phi}(1+\cos\beta)^2+(1-\cos\beta)^2+2\sin^2\beta\tan^2\phi-2\frac{\sqrt{\cos2\phi}}{\cos\phi}\sin^2\beta\cos2\alpha}{8}+$$
$$+\frac{\frac{\sqrt{\cos2\phi}\sin\phi}{\cos^2\phi}\sin\beta(1+\cos\beta)\cos\alpha-\tan\phi\sin\beta(1-\cos\beta)\cos\alpha}{2\sqrt{2}};$$ $$\omega_5(0,\overrightarrow{n})=
\frac{2\cos^2\beta\tan^2\phi+\sin^2\beta+\frac{\cos2\phi}{\cos^2\phi}\sin^2\beta+2\frac{\sqrt{\cos2\phi}}{\cos\phi}\sin^2\beta\cos2\alpha}{4}-\frac {\tan\phi\sin2\beta\cos\alpha}{2\sqrt{2}}-$$
$$-\frac{\frac{\sqrt{\cos2\phi}}{\cos^2\phi}\sin\phi\sin2\beta\cos\alpha}{2\sqrt{2}};$$
$$\omega_5(-1,\overrightarrow{n})=
\frac{\frac{\cos2\phi}{\cos^2\phi}(1-\cos\beta)^2+(1+\cos\beta)^2+2\sin^2\beta\tan^2\phi-2\frac{\sqrt{\cos2\phi}}{\cos\phi}\sin^2\beta\cos2\alpha}{8}-$$
$$-\frac{\frac{\sqrt{\cos2\phi}\sin\phi}{\cos^2\phi}\sin\beta(1-\cos\beta)\cos\alpha-\tan\phi\sin\beta(1+\cos\beta)\cos\alpha}{2\sqrt{2}}.$$
The spin tomogram for $|\psi\rangle$ was presented in section 4.


\begin{thebibliography}{0}

\bibitem{1}
J.S. Bell, Physics \textbf{1}, 195 (1964).

\bibitem{2}
S.Kochen and E.P. Specker, J. Math. Mech. \textbf{17}, 59 (1967).

\bibitem{3}
A.A. Klyachko, M.A. Can, S. Binicio$\breve{\text{g}}$lu and A.S. Shumovsky, Phys. Rev. Lett. \textbf{101}, 020403 (2008).

\bibitem{4}
P. Kurzynski, R. Ramanathan and D. Kaszlikowski, Phys. Rev. Lett. \textbf{109}, 020404 (2012).

\bibitem{5}
O. G$\ddot{\text{u}}$hne, C. Budroni, A. Cabello, M. Kleinmann, J.-A. Larsson, arXiv/quant-ph: 1302.2266 (2013).

\bibitem{6}
J. Ahrens, E. Amselem, A. Cabello and M. Bourennane, arXiv/quant-ph: 1301.2887v2 (2013).

\bibitem{7}
A.E. Rastegin, Quantum Information Processing \textbf{11}, 1895-1910 (2012).

\bibitem{8}
A. Cabello, M.T. Cunha, Phys. Rev. A \textbf{87}, 022126 (2013).

\bibitem{9}
M. Markiewicz, P. Kurzynski, J. Thompson, S.-Y. Lee, A. Soeda, T. Paterek, D. Kaszlikowski, Unified approach to contextuality, non-locality, and temporal correlations, arXiv/quant-ph: 1302.3502 (2013).

\bibitem{10}
S. Mancini, V.I. Man'ko and P. Tombesi, Quantum Semiclass. Opt. \textbf{7}, 615 (1995).

\bibitem{11}
A. Ibort, V.I. Man'ko, G. Marmo, A. Simoni, F. Ventriglia, Phys. Scr. \textbf{79}, 065013 (2009).

\bibitem{12}
A. Cabello, Phys. Rev. Lett. \textbf{101}, 210401 (2008).

\bibitem{13}
A. Peres, Phys. Lett. A \textbf{151}, 107 (1990).

\bibitem{14}
N.D. Mermin, Phys. Rev. Lett. \textbf{65}, 3373 (1990).

\bibitem{15}
V.I. Man'ko and O.V. Man'ko, J. Exp. Theor. Phys. \textbf{85}, 430 (1997).

\bibitem{16}
S.N. Filippov, V.I. Man'ko, Purity of spin states in terms of tomograms, Journal of Russian Laser Research (2013).

\bibitem{17}
V.I. Man'ko, G. Marmo, E.C.G. Sudarshan, F. Zaccaria, arXiv/quant-ph: 0705.3574 (2007).

\bibitem{18}
P. Dit$\breve{\text{a}}$, Factorization of Unitary Matrices, arXiv/math-ph: 0103005 (2001).

\end{thebibliography}
\end{document}